\documentclass{article}
\usepackage{arxiv}
\usepackage[utf8]{inputenc}
\usepackage[T1]{fontenc}
\usepackage{xcolor}
\usepackage[sort&compress, round]{natbib}
\definecolor{blendedblue}{rgb}{0.2, 0.2, 0.6}
\usepackage{graphicx}
\usepackage{enumerate}
\usepackage[bookmarks   = true,
            citecolor   = blendedblue,
            colorlinks  = true,
            linkcolor   = blendedblue,
            urlcolor    = blendedblue,
            citecolor   = blendedblue,
            linktocpage = false,
            hyperindex  = true]{hyperref}
\usepackage{booktabs}
\usepackage{amsfonts}
\usepackage{nicefrac}
\usepackage{amsmath, amssymb, amsthm}
\usepackage{url}
\usepackage{doi}
\usepackage{bm}
\usepackage{todonotes}
\usepackage{fontawesome}
\usepackage{mathtools}
\usepackage{lmodern}
\usepackage{siunitx}
\usepackage{booktabs}
\usepackage{etoolbox}
\usepackage{calc}
\usepackage{enumitem}
\usepackage{dsfont}
\usepackage{multirow}
\usepackage{array}
\usepackage{graphicx}
\definecolor{blendedblue}{rgb}{0.2, 0.2, 0.6}

\newcommand{\E}{{\mathbb{E}}}
\newcommand{\Cov}{{\operatorname{Cov}}}

\usepackage{enumerate} 
\usepackage{orcidlink}

\usepackage[labelsep=period]{caption}

\renewcommand{\headeright}{}

\usepackage{parskip}
\makeatletter 
\renewcommand\@biblabel[1]{#1.}
\makeatother

\title{Multivariate Long-term Profile Monitoring with Application to the KW51 Railway Bridge}
\renewcommand{\shorttitle}{Multivariate CAFDA for SHM}
\date{}

\author{Philipp Wittenberg\orcidlink{0000-0001-7151-8243}\\
 	Dept. of Mathematics and Statistics\\
        Institute of Artificial Intelligence\\
	School of Economics and Social Sciences\\
        Helmut Schmidt University\\
	Hamburg, Germany\\
	\texttt{pwitten@hsu-hh.de} \\
	\And
        Alexander Mendler\orcidlink{0000-0002-7492-6194}\\
        Dept. of Materials Engineering\\
        TUM School of Engineering and Design\\
	Technical University of Munich\\
        Munich, Germany\\
	\texttt{alexander.mendler@tum.de}\\
        \And 
        Sven Knoth\orcidlink{0000-0002-9666-5554}\\
 	Dept. of Mathematics and Statistics\\
 	Institute of Artificial Intelligence\\
        School of Economics and Social Sciences\\
        Helmut Schmidt University\\
	Hamburg, Germany\\
	\texttt{knoth@hsu-hh.de} \\
	\And 
        Jan Gertheiss\orcidlink{0000-0001-6777-4746}\\
 	Dept. of Mathematics and Statistics\\
        Institute of Artificial Intelligence\\
 	School of Economics and Social Sciences\\
        Helmut Schmidt University\\
	Hamburg, Germany\\
	\texttt{gertheij@hsu-hh.de} \\
}

\begin{document}	
\maketitle

\begin{abstract}
Structural Health Monitoring (SHM) plays a pivotal role in modern civil engineering, providing critical insights into the health and integrity of infrastructure systems. This work presents a novel multivariate long-term profile monitoring approach to eliminate fluctuations in the measured response quantities, e.g., caused by environmental influences or measurement error. Our methodology addresses critical challenges in SHM and combines supervised methods with unsupervised, principal component analysis-based approaches in a single overarching framework, offering both flexibility and robustness in handling real-world large and/or sparse sensor data streams. We propose a function-on-function regression framework, which leverages functional data analysis for multivariate sensor data and integrates nonlinear modeling techniques, mitigating covariate-induced variations that can obscure structural changes. 
\end{abstract}

\vfill

\section{Introduction and Data}\label{sec:introdata}
Structural Health Monitoring (SHM) is a critical component in ensuring the safety and reliability of infrastructure. SHM often uses statistical tools like principal component analysis or different types of (multivariate) regression techniques to analyze and address fluctuations in vibration-based properties, such as natural frequency data~\citep{Hu2017}. This paper presents a new multivariate monitoring approach based on Functional Data Analysis (FDA). While FDA offers a robust framework for examining complex, high-dimensional multivariate sensor data, it also integrates nonlinear modeling techniques, mitigating covariate-induced variations that can obscure structural changes. However, so far, its application in SHM has been limited~\citep{Wittenberg.etal_2025,Wittenberg.etal_2024b}.

\begin{figure}[!htb]
\includegraphics[width=1\textwidth]{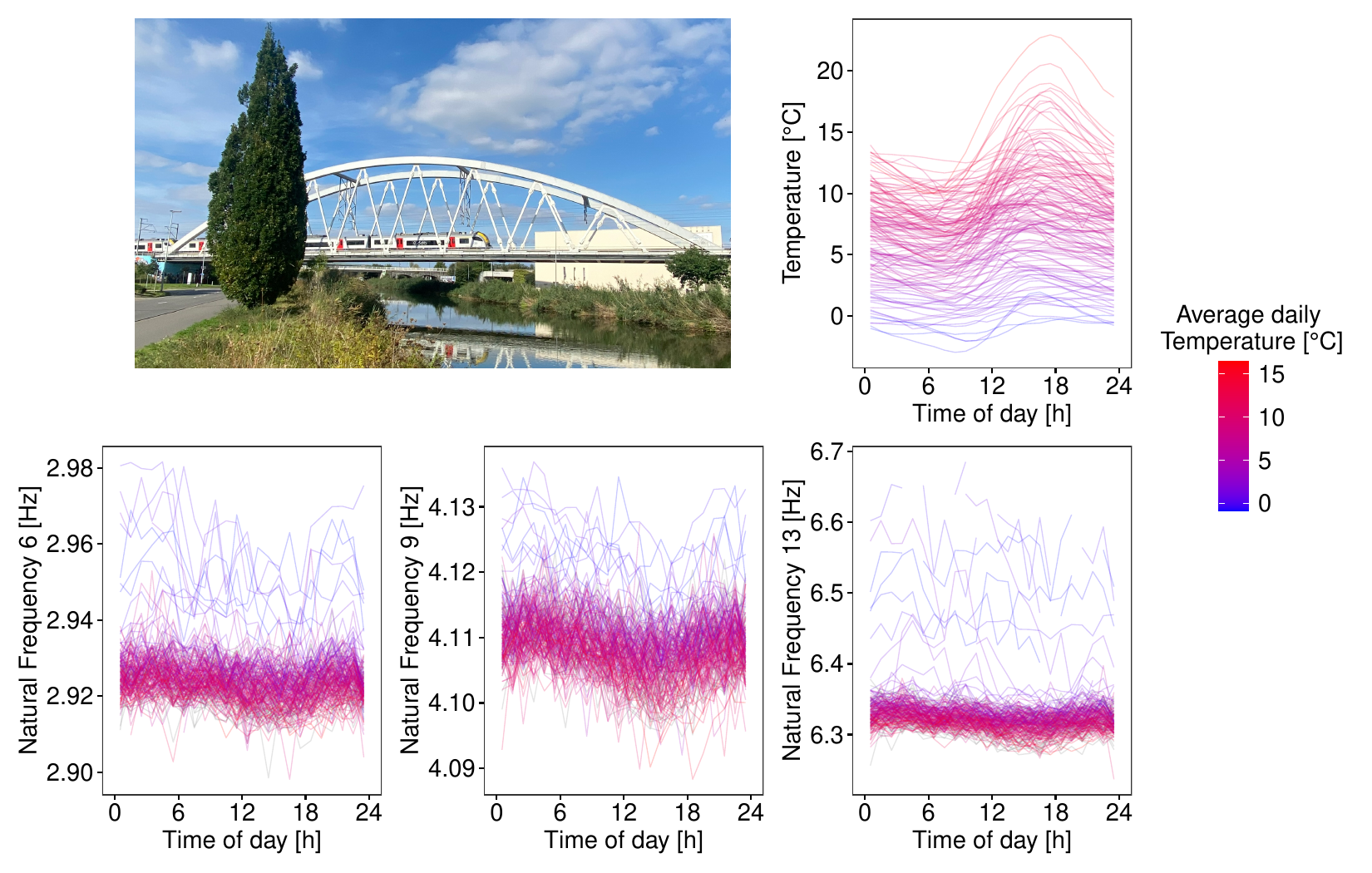}
\caption{KW51 railway bridge in September 2024 (top-left), daily profiles of steel temperature (top-right), and natural frequencies of modes six, nine, and thirteen (bottom row) with a one-hour sampling rate of the first 200 days. The profiles are highlighted in color according to their average daily temperature.} 
\label{fig:KW51_photo_profiles}
\end{figure}

We will illustrate our methodology using recently published vibration-based natural frequency data for the KW51 railroad bridge~\citep{Maes.Lombaert_2020}. The bridge spans 115 meters in length and 12.4 meters in width, situated between Leuven and Brussels, Belgium, along the L36N railway line. It was monitored from October 2, 2018, to January 15, 2020, with a retrofitting period from May 15 to September 27, 2019. Various quantities, such as the steel surface temperature, were measured hourly \citep{Maes.Lombaert_2020,Maes.Lombaert_2021,Maes.etal_2022}. Here, we will focus on the data for three modes, 6, 9, and 13, where some extreme outliers corresponding to some data points resulting from abnormal bridge behavior on particularly cold days were removed from the data set. Additionally, we consider the potentially confounding variable, temperature at the bridge deck level measured directly at the site \citep{Maes.Lombaert_2021,Maes.etal_2022}.
Importantly, we do not consider single measurement points but 24h profiles, i.e., functions as shown in Figure~\ref{fig:KW51_photo_profiles}, as the quantities of interest. This functional perspective has, among other things, the advantage that recurring (daily or yearly) patterns, as well as correlated errors, can be taken into account, and this additional information can be exploited to improve damage detection~\citep{Wittenberg.etal_2025}. 

The rest of the paper is structured as follows. Section~\ref{sec:methods} describes our methods that exploit the functional nature of the data. Section~\ref{sec:results} shows the results for the KW51 data, and Section~\ref{sec:discussion_conclusion} concludes.
\section{Methods}\label{sec:methods}
The model we assume for ``in-control'' (IC) data (Phase-I) has the following basic form. To keep things simple, we restrict ourselves to a single, functional covariate $z_j(t)$, e.g., denoting the temperature at time $t \in \mathcal{T}, \mathcal{T} = (0\text{h}, 24\text{h}]$, and day $j$, but consider multiple system outputs $u_{j,q}(t)$, $q = 1,\ldots,Q$. The latter could be raw sensor measurements (yet preprocessed to some extent), such as strain or inclination data, or extracted features, such as natural frequencies. Then, we assume the basic model
\begin{equation}\label{eq:basic_model}
 u_{j,q}(t) = \alpha_q(t) + f_q(z_j(t)) + E_{j,q}(t), 
\end{equation}
where $\alpha_q(t)$ is a fixed functional intercept, $f_q(z_j(t))$ is a fixed, potentially non-linear effect of temperature, and $E_{j,q}(t)$ is a day-specific, functional error term with zero mean and a common covariance across days, i.e., $\E(E_{j,q}(t)) = 0$, $\Cov(E_{j,q}(s), E_{j,q}(t)) = \Sigma_q(s,t)$, $s,t \in \mathcal{T}$. In the FDA framework used here, sampling instances are days instead of single measurement points, and the daily profiles are considered the quantities of interest. This model has several advantages over scalar-on-scalar(s) regression as typically used for response surface modeling in SHM \citep{Wittenberg.etal_2024b}: First, the functional intercepts $\alpha_q(t)$ capture recurring daily patterns that cannot be explained through the available environmental or operational variables, e.g., because the factors causing them are not recorded/available. Secondly, the error terms $E_{j,q}(t)$ are typically not white noise but correlated over time, i.e., in the $t$-direction. Also, variances may vary over the day. For instance, error variances may be lower at night due to the lower traffic volume and no changes in the solar radiation. In other words, $\Sigma_q(s,t)$ is not necessarily zero for $s \neq t$, and $\Sigma_q(t,t)$ is not constant. Furhermore, in the case of SHM, there is another important aspect to consider with respect to $E_{j,q}(t)$: Those processes contain the relevant information for the monitoring task since they capture deviations from the system outputs $\alpha_q(t) + f_q(z_j(t))$ that would be expected for a specific, let us say, temperature at time $t$ if the structure is ``in-control''. For exploiting this information, we decompose each process $E_{j,q}(t)$ into a more structural component $w_{j,q}(t)$ and white noise $\epsilon_{j,q}(t)$ with variance $\sigma_q^2$ in terms of
\begin{equation}\label{eq:Ej1}
E_{j,q}(t) = w_{j,q}(t) + \epsilon_{j,q}(t).
\end{equation}
Since $\epsilon_{j,q}(t)$ is assumed to be pure noise, it does not carry relevant information, and $w_{j,q}(t)$ should be the part to focus on for monitoring purposes. The latter is modeled analogously to~\cite{Wittenberg.etal_2025} through 
\begin{equation}\label{eq:wj}
w_{j,q}(t) = \sum_{r=1}^{m_q} \xi_{j,q,r}\phi_{q,r}(t),
\end{equation}
where $\phi_{q,r}$, $r=1,\ldots,m_q$, are the first $m_q$ (orthonormal) eigenfunctions of the covariance $\Sigma_q(s,t)$, obtained through functional principal component analysis (FPCA)~\citep{Yao.etal_2005}. A common approach to choose an appropriate $m_q$ is to ensure that at least 95\% or 99\% of the overall variance is explained~\citep{Gertheiss.etal_2024}, and we use 99\% throughout the paper. The so-called scores $\xi_{j,q,r}$, $r=1,\ldots,m_q$, $q=1,\ldots,Q$, can then be used as damage-sensitive features for monitoring. As the functions $\alpha_q, f_q, \phi_{q,r}$ are estimated from IC data in the model training phase, it is the scores obtained for future data that tell us whether the system outputs deviate from the values that would be expected for an IC structure over the day for given values of the covariate.

For estimating the functions $\alpha_q$ and $f_q$ from \eqref{eq:basic_model}, we follow the same semiparametric approach as used in \cite{Wittenberg.etal_2025}. The unknown function, say $f_q$, is expanded in basis functions such that (compare, e.g., \cite{Greven.Scheipl_2017} and \cite{Wood_2017})
\begin{equation}\label{eq:f_basis}
    f_q(z) = \sum_{l=1}^{L_q} \gamma_{ql}b_{ql}(z).
\end{equation}
A popular choice for $b_{q1}(z),\ldots,b_{qL_q}(z)$ is a cubic B-spline basis, which means that $f_q$ is a cubic spline function \citep{deBoor_1978,Dierckx_1993}. For being sufficiently flexible with respect to the types of functions that can be fitted through \eqref{eq:f_basis}, typically, a rich basis with a large $L_q$ is chosen, and we use $L_q = 10$ for all $q$ throughout the paper. A large $L_q$, however, often leads to wiggly estimated functions if the basis coefficients $\gamma_{q1},\ldots, \gamma_{qL_q}$ are fit without any smoothness constraint. The latter is typically imposed by adding a so-called \emph{penalty} term when fitting the unknown coefficients through least-squares or maximum likelihood. A popular penalty is the integrated squared second derivative $\int_{\mathcal{D}_q} [f_q''(z)]^2 dz$, where $\mathcal{D}_q$ is the domain of $f_q$. Since we focus on the confounder temperature here, $\mathcal{D}_q = \mathcal{D}$ is the same for each $q$. If the eigenfunctions $\phi_{q,1},\ldots,\phi_{q,m_q}$ are known, model \eqref{eq:basic_model} is an additive mixed model with random effects $\xi_{j,q,r}$, which can be estimated using R add-on package \texttt{mgcv}~\citep{Wood_2017}. Hence, again analogously to \cite{Wittenberg.etal_2025}, we use a two-step approach: We first fit model~\eqref{eq:basic_model} with a working independence assumption concerning $E_{j,q}(t)$ \citep{Scheipl.etal_2015}. Then, we use the resulting estimates of the error process for FPCA, plug in the estimated eigenfunctions $\hat{\phi}_{r,q}$, and fit the final model.\\

Once the functional mixed model~\eqref{eq:basic_model} has been trained for each $q=1,\ldots,Q$, it can be used to monitor future system outputs if the covariate in the trained models is available as well. An essential input for monitoring are the principal component scores $\xi_{g,q,1},\ldots,\xi_{g,q,m_q}$ for a new day $g$ in the online monitoring data (Phase-II). To estimate those scores from the new data, we first use the fixed effects from the Phase-I model, i.e., $\hat{\alpha}_q(t_{gi})$ and $\hat{f}_q(z_g(t_{gi}))$, to obtain a prediction for the system outputs $u_{g,q}(t_{gi})$. Here, $t_{gi}$, $i=1,\ldots,N_g$, denote the time points observations of the covariate $z$ are available on day $g$. Then, those predictions are subtracted from the actually observed $u_{g,q}(t_{gi})$ to obtain estimated measurements $\hat{E}_{g,q}(t_{gi})$ of the error process. 
For estimating the scores, we employ the following approach that uses the interpretation of \eqref{eq:basic_model} as a mixed model: Let $\boldsymbol{\phi}_{g,q,r} = (\phi_{q,r}(t_{g1}),\ldots,\phi_{q,r}(t_{g N_g}))^\top$ be the $r$th eigenfunction of the $q$th system output evaluated at time points $t_{gi}$, $i=1,\ldots, N_g$, $r=1,\ldots,m_q$, and $\Sigma_{\mathbf{E}_{g,q}}$ the covariance matrix of the error vector $\mathbf{E}_{g,q} = (E_{g,q}(t_{g1}),\ldots, E_{g,q}(t_{g N_g}))^\top$. Then, assuming a Gaussian distribution, the conditional expectation of the score $\xi_{g,q,r}$ given $\mathbf{E}_{g,q}$ is  
$\mathbb{E}(\xi_{g,q,r}|\mathbf{E}_{g,q}) = \nu_{q,r}^2 \boldsymbol{\phi}_{g,q,r}^\top \Sigma_{\mathbf{E}_{g,q}}^{-1}\mathbf{E}_{g,q}, \;\; r=1,\ldots,m_q$ \citep{Yao.etal_2005}; $\nu_{q,r}^2$ is the variance of the $r$th score of output $q$, which is a so-called variance component in the mixed model framework and can be estimated using (restricted) maximum likelihood (RE)ML. Due to \eqref{eq:Ej1} and \eqref{eq:wj}, the matrix $\Sigma_{\mathbf{E}_{g,q}}$ can be estimated as $\hat{\Sigma}_{\mathbf{E}_{g,q}} = \hat{\boldsymbol{\Phi}}_{g,q} \mbox{diag}(\hat{\nu}_{q,1}^2,\ldots,\hat{\nu}_{q,m_q}^2) \hat{\boldsymbol{\Phi}}_{g,q}^\top + \hat{\sigma}_q^2\mathbf{I}_{N_g}$, with $\hat{\boldsymbol{\Phi}}_{g,q} = (\hat{\boldsymbol{\phi}}_{g,q,1}|\ldots|\hat{\boldsymbol{\phi}}_{g,q,m_q})$. After plugging in the estimates $\hat{\sigma}_q^2$, $\hat{\nu}_{q,r}^2$, $\hat{\phi}_{q,r}$, $r=1,\ldots,m_q$, from the model training phase and $\hat{\mathbf{E}}_{g,q} = (\hat{E}_{g,q}(t_{g1}),\ldots,\hat{E}_{g,q}(t_{g N_g}))^\top$ from above, we obtain the estimated scores
\begin{equation}\label{eq:estSc2}
    \hat{\xi}_{g,q,r} = \hat{\nu}_{q,r}^2 \hat{\boldsymbol{\phi}}_{g,q,r}^\top \hat{\Sigma}_{\mathbf{E}_{g,q}}^{-1}\hat{\mathbf{E}}_{g,q}, \;\; r=1,\ldots,m_q, \; q = 1,\ldots,Q.
\end{equation}

These scores can then be used as input to a control chart for monitoring purposes. For instance, a multivariate Hotelling control chart is often used to detect a shift in the structural condition \citep{Deraemaeker.etal_2008,Magalhaes.etal_2012,Comanducci.etal_2016}. Here, we employ a memory-type control chart, the Multivariate Exponentially Weighted Moving Average (MEWMA)
\citep{Lowry.etal_1992} to the Phase-II scores $\hat{\xi}_{g,q,r}$, $g=1,2,\ldots$ from above. The MEWMA chart contains the Hotelling chart as a special case.
For both charts, one assumes serial independent normally distributed vectors $\hat{\boldsymbol{\xi}}_1, \hat{\boldsymbol{\xi}}_2, \dots$ with $\hat{\boldsymbol{\xi}}_g\sim\mathcal{N}(\boldsymbol{\mu}, \boldsymbol{\Lambda})$
where the scores $\hat{\xi}_{g,q,r}$ from \eqref{eq:estSc2} are collected in $\boldsymbol{\xi}_g$ in terms of $\hat{\boldsymbol{\xi}}_g = (\hat{\boldsymbol{\xi}}_{g,1}^\top,\ldots,\hat{\boldsymbol{\xi}}_{g,Q}^\top)^\top$, with $\hat{\boldsymbol{\xi}}_{g,q} = (\hat{\xi}_{g,q,1},\ldots,\hat{\xi}_{g,q,m_q})^\top$. Follow the notation in \cite{Knoth_2017}, we define a mean vector $\boldsymbol{\mu}$ that adheres to the change point model $\boldsymbol{\mu}=\boldsymbol{\mu}_0$ for $g<\tau$ and $\boldsymbol{\mu}=\boldsymbol{\mu}_1$ for $g\geq\tau$ for an unknown time point $\tau$ and by definition $\boldsymbol{\mu}_0=\boldsymbol{0}$. 
Here, we estimate the covariance matrix $\boldsymbol{\Lambda}$ based on the scores $\hat{\xi}_{g,q,r}$ from the in-control data.
Then, we apply the following smoothing procedure to compute the MEWMA statistic
\begin{equation}\label{eq:MEWMA}
 \boldsymbol{\omega}_g=(1-\lambda)\boldsymbol{\omega}_{g-1} + \lambda\boldsymbol{\xi}_g, \quad  \boldsymbol{\omega}_0=\boldsymbol{0}\
\end{equation}
with $g=1, 2, \dots$ and smoothing constant $0 < \lambda \leq 1$. The latter parameter $\lambda$ controls the sensitivity to the shift to be detected. Smaller values of $\lambda$ such as $\lambda \in \{0.1, 0.2, 0.3\}$ are typically deployed to detect smaller shifts
(refer to \textit{``In general, values for $r$ from 0.1 to 0.5 are good choices''} in \cite{Prab:Rung:1997}, where $r$ denoted the smoothing constant), while $\lambda=1$ results in the Hotelling chart. In this study, we use $\lambda=0.3$ and $\lambda = 1$. The control statistic is the Mahalanobis distance
\begin{equation}\label{eq:MHD}
T^2_g=(\boldsymbol{\omega}_g-\boldsymbol{\mu}_0)^\top\boldsymbol{\Lambda}^{-1}_{\boldsymbol{\omega}}   (\boldsymbol{\omega}_g-\boldsymbol{\mu}_0),
\end{equation}
with asymptotic covariance matrix of $\boldsymbol{\omega}_g$,   $\boldsymbol{\Lambda}_{\boldsymbol{\omega}}=\lim_{g\rightarrow\infty} \text{Cov}(\boldsymbol{\omega}_g) = \Big\{\frac{\lambda}{2-\lambda}\Big\}\boldsymbol{\Lambda}$.

The MEWMA chart issues an alarm if $T_i^2>h_4$, i.e., the control statistic is above the threshold value $h_4$. The expected value of the stopping time $N=\min\big\{g \ge 1\!: T^2_g> h_4\big\}$, also known as (zero-state) average run length (ARL), is often used to measure the control chart's performance. It is defined as the average number of observations until the chart signals an alarm. If the process is in control, the ARL (ARL$_0$) should be high to avoid false alarms. If there is a change in the underlying process, the ARL (ARL$_1$) should be low to detect changes quickly. To determine the threshold value, the ARL must be calculated when the process is in control, usually applying a grid search or a secant rule. This ARL$_0$ can be calculated as described in \cite{Knoth_2017} and is implemented in R-package \texttt{spc} \citep{Knoth_2024}.
\section{Application to SHM data}\label{sec:results}
This section applies the proposed multivariate, FDA-based monitoring approach to the natural frequencies of three different modes of the KW51 bridge shown in Figure \ref{sec:introdata}. This includes the retrofitting data for online monitoring, as detailed in \cite{Maes.etal_2022}. We use two configurations of the MEWMA chart with different values of $\lambda \in \{0.3,1\}$. Phase-I consists of the first 200 days; however, due to variations in data availability for both the response and covariate variables, we can only use between 145 and 148 profiles for individual modeling. During this period, the overall proportion of missing data points is high, reaching 47.5\%. We start with the model presented in \eqref{eq:basic_model} for each natural frequency. Each model includes a functional intercept $\alpha_q(t)$, a potentially nonlinear temperature effect $f_q(z(t))$, and the structural component of the error process $E_{j,q}(t)$. As sketched in Section~\ref{sec:methods}, after estimating the fixed effects from the initial models, we apply FPCA to the residual profiles which contain a sufficient number of data points from both the frequency and temperature curves. Subsequently, the models are refitted by incorporating the eigenfunctions to account for the functional random effects. More details about this covariate-adjusted functional data analysis can be found in \cite{Wittenberg.etal_2025}.

\begin{figure}[!htb]
\includegraphics[width=1\textwidth]{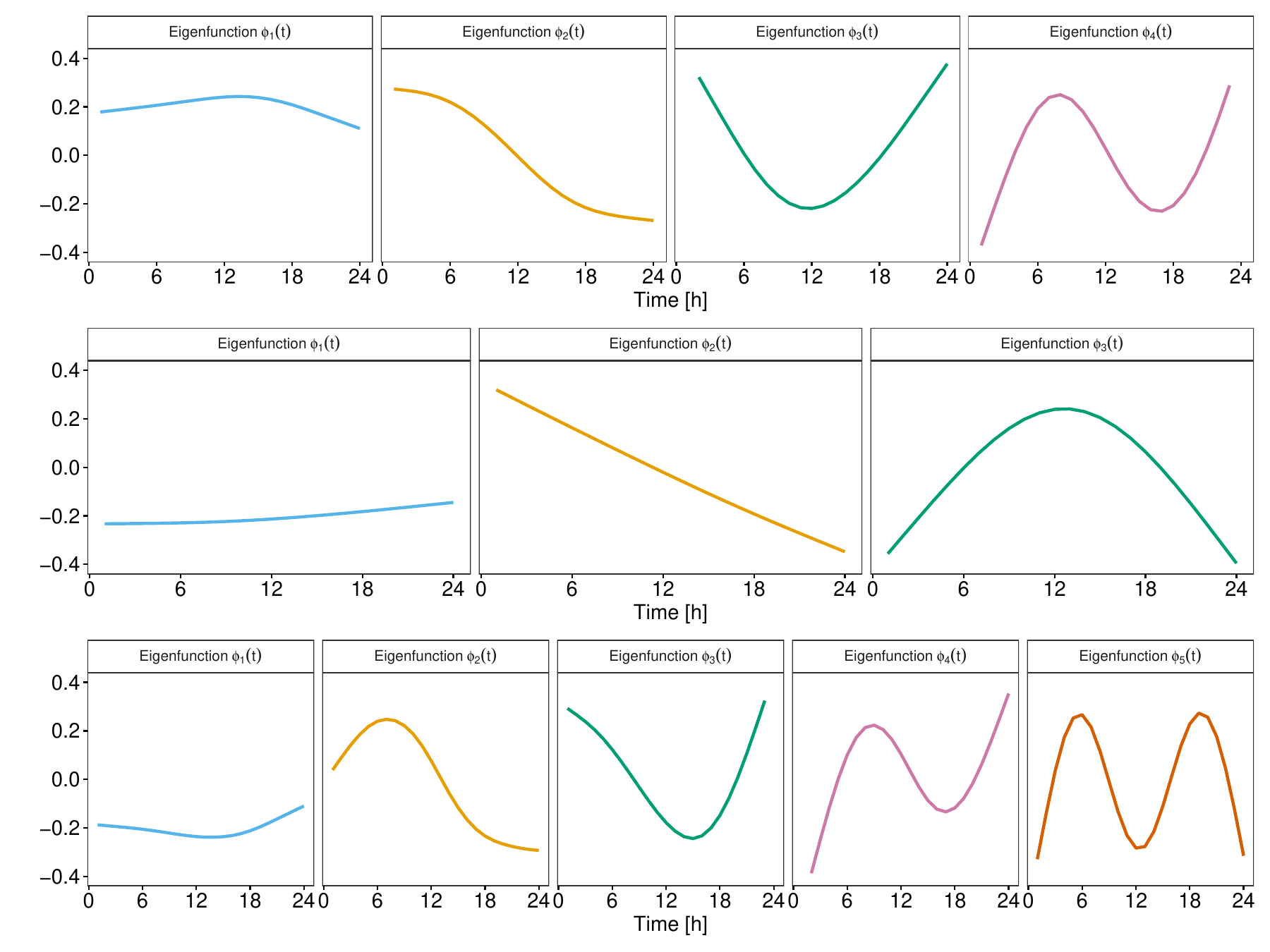}
\caption{Estimated eigenfunctions $\phi_{q,r}$ of the natural frequency of mode~6~($q=1$, top), mode~9~($q=2$, middle), and 13~($q=3$, bottom), $r = 1,\ldots, m_q$, with $m_1 = 4$, $m_2 = 3$, $m_3 = 5$.} \label{fig:eigenfunctions}
\end{figure}

Figure \ref{fig:eigenfunctions} displays the eigenfunctions $\hat{\phi}_{q,1}(t), \ldots, \hat{\phi}_{q,m_q}(t)$ for the natural frequencies of the three modes under consideration, i.e., $q=1,2,3$. Note that those eigenfunctions are only identifiable up to the sign. Further note that $m_q$, which indicates the number of eigenfunctions required to account for 99\% of the variance in the structural component $w_{j,q}$ \eqref{eq:wj} of the error process, differs among the modes, with values of $m_1=4$, $m_2=3$, and $m_3=5$. 
Despite this variation, similar shapes can be observed across the eigenfunctions of the natural frequencies. For instance, the first three eigenfunctions of all modes exhibit qualitatively similar shapes. Additionally, $\phi_{1,4}$ and $\phi_{3,4}$ are also comparable. As detailed further in \cite{Wittenberg.etal_2025}, these eigenfunctions possess straightforward and interpretable shapes. For example, the first eigenfunction closely resembles a horizontal line, indicating that the first principal component represents a (weighted) average of the daily errors, with maximum weights occurring in the early afternoon for both $\phi_{1,1}$ and $\phi_{3,1}$. The second component reflects the differences in errors between morning and evening hours, while the third component captures the contrast between night and day.

\begin{figure}[!htb]
    \centering
    \includegraphics[width = 1\textwidth]{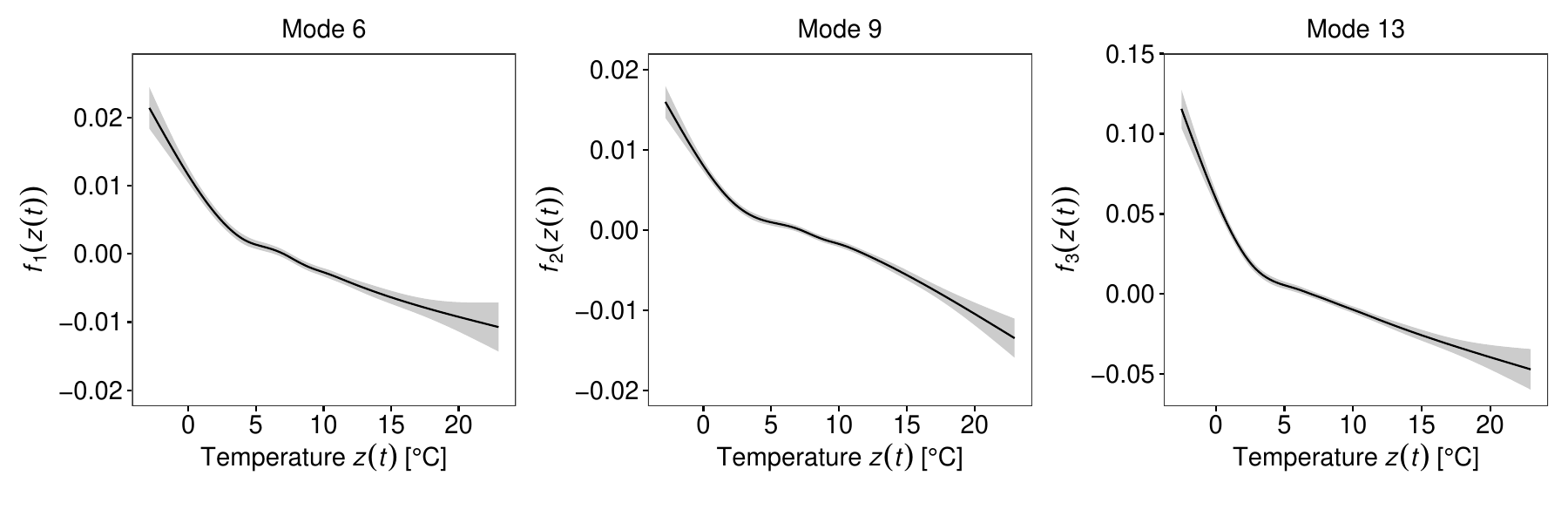}
    \caption{Results of the functional modeling approach for the nonlinear effect of temperature $f_q(z(t))$ on the natural frequency of modes 6 ($q=1$, left), 9 ($q=2$, middle), and 13 ($q=3$, right) of the KW51 bridge.}
    \label{fig:effect_plots}
\end{figure}

Figure \ref{fig:effect_plots} illustrates the centered nonlinear fixed effects $f_q(z)$ of temperature $z$ on the natural frequency for each of the three modes separately. The grey-shaded areas represent the uncertainty of the estimated effects, depicted as pointwise 95\% confidence intervals. While the centered functional intercept $\alpha_q(t)$ (not shown), which represents a recurring daily pattern, remains fairly constant throughout the day across all three modes and, thus, shows minimal influence on the natural frequencies, the temperature effects presented in Figure \ref{fig:effect_plots} all display distinct nonlinear shapes. Notably, there is a kink occurring between 2°C and 3°C for all modes. This effect size is more pronounced in absolute terms for mode 13, as evidenced by the approximately tenfold larger scale on the Y-axis. However, it is important to note the wider range of data variation for these frequencies (see Figure \ref{fig:effect_plots}). In summary, the estimated fixed effects are consistent with existing literature, which suggests that the impact of temperature on dynamic response follows an inverse relationship, being more pronounced at lower temperatures \citep{Xia.etal_2012,Han.etal_2021}. The overall model fits, represented by \(R^2\) values taking only the fixed effects into account, are 0.43 for mode 6, 0.46 for mode 9, and 0.56 for mode 13, respectively, indicating that around 50\% of the overall variation is explained by changes in the temperature.

\begin{figure}[!htb]
\includegraphics[width=1\textwidth]{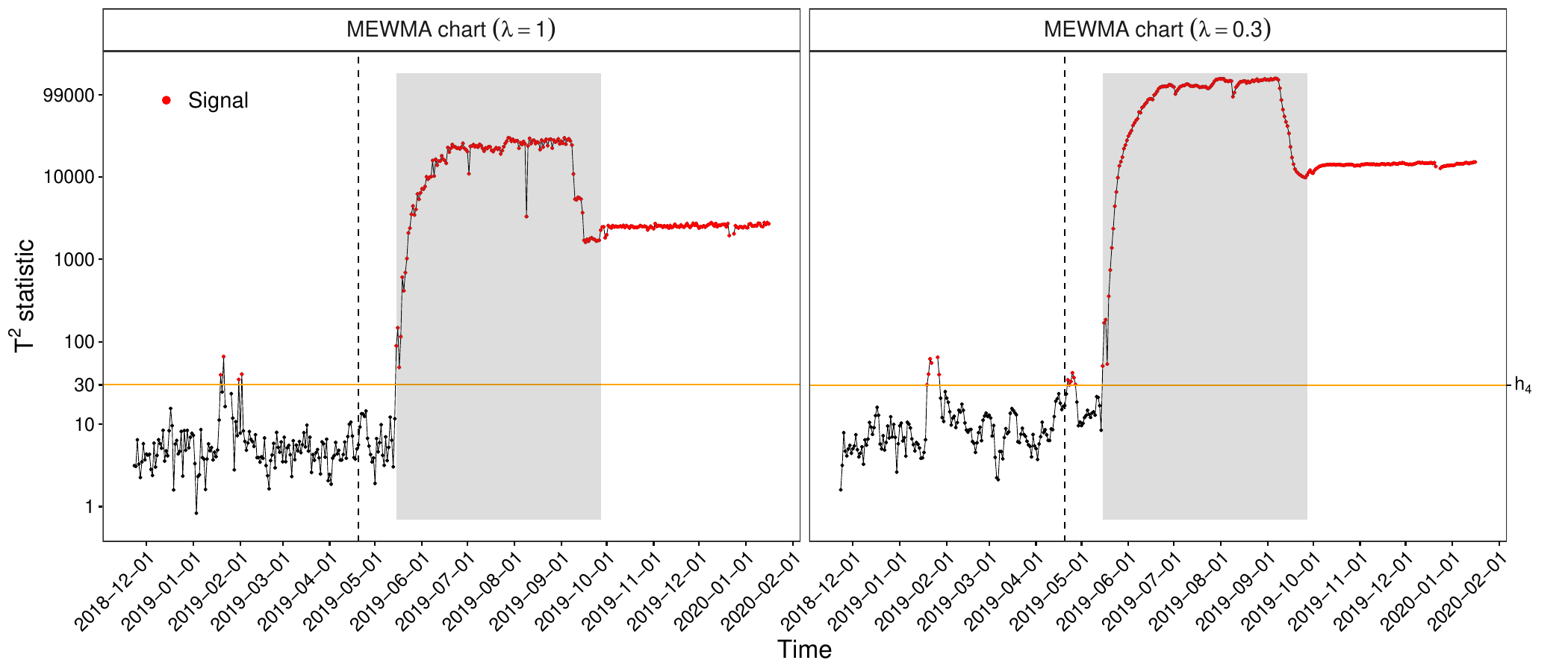}
\caption{MEWMA control charts for natural frequencies 6, 9 and 13 with smoothing parameter $\lambda \in \{1,0.3\}$ on a logarithmic y-axis, using the functional model for the KW51 bridge including the retrofitting period (grey shaded area).}
\label{fig:MEMWA_v1}
\end{figure}

The MEWMA control chart presented in Section \ref{sec:methods} is applied in two different configurations for comparison to detect potential anomalies as a final step in our multivariate functional covariate-adjusted monitoring procedure. Both control charts are calibrated to an ARL$_0 = 370.4$ using thresholds $h_4=30.09$ and $h_4=29.64$ for $\lambda=1$ and $\lambda=0.3$, respectively. Utilizing the estimated parameters and the incoming data, the scores \eqref{eq:estSc2} of the 12 principal components from Fig.~\ref{fig:eigenfunctions} are used in \eqref{eq:MHD} to calculate the control statistic $T^2_g$ for day $g$ (compare Section~\ref{sec:methods}). Figure \ref{fig:MEMWA_v1} shows the resulting control charts, where the dashed vertical line indicates the end of the in-control Phase and the start of the online monitoring. It is evident that both charts immediately signal alarms for an out-of-control situation at the beginning of the retrofitting phase and reach a higher level at the end, thus effectively detecting a sustained shift in the data or a change in the bridge's dynamic response. However, we also note several false alarms: 12 false alarms for $\lambda=0.3$ (occurring in two clusters from January 19, 2019, to January 27, 2019, and from April 22, 2020, to April 27, 2020), and four false alarms for $\lambda=1$ (from January 19, 2019, to February 2, 2019). These false alarms may be related to cold temperatures at the boundary of the domain of the regression functions shown in Fig.~\ref{fig:effect_plots}. In summary, however, the presented method provided a flexible and interpretable modeling of the influence of the nonlinear covariate temperature and a rapid detection of the retrofitting phase at the bridge.

\section{Conclusion and outlook}\label{sec:discussion_conclusion}
This paper presented a novel multivariate monitoring approach for SHM based on FDA, extending a recently introduced framework \citep{Wittenberg.etal_2025}. The proposed method explicitly leverages the functional nature of the data. Hence, it offers the flexibility to model recurring daily and yearly patterns alongside environmental or operational influences. Our approach flexibly handles diverse data types, including missing observations and sparse and aggregated to high-resolution, dense data. Furthermore, utilizing state-of-the-art functional additive mixed models facilitates nonlinear modeling. In a real-world application to the KW51 bridge data, we demonstrated that the estimated eigenfunctions of the natural frequencies exhibit similar shapes across modes, suggesting that the first few principal components capture the dominant patterns in the data. The estimated fixed effects of temperature on the natural frequencies displayed distinct nonlinear shapes. Combining this adaptable and interpretable functional modeling with an advanced memory-type MEWMA control chart to detect potential anomalies in the data provided a reliable basis for decision-making. Future work will focus on refining the method to address issues of false alarms and exploring its application to other SHM datasets to assess its generalizability and robustness. 
\section*{Acknowledgements}
This research paper out of the project `SHM -- Digitalisierung und Überwachung von Infrastrukturbauwerken' is funded by dtec.bw -- Digitalization and Technology Research Center of the Bundeswehr, which we gratefully acknowledge. dtec.bw is funded by the European Union -- NextGenerationEU.

\bibliographystyle{abbrvnat}

\begin{thebibliography}{22}
\providecommand{\natexlab}[1]{#1}
\providecommand{\url}[1]{\texttt{#1}}
\expandafter\ifx\csname urlstyle\endcsname\relax
  \providecommand{\doi}[1]{doi: #1}\else
  \providecommand{\doi}{doi: \begingroup \urlstyle{rm}\Url}\fi

\bibitem[de~Boor(1978)]{deBoor_1978}
C.~de~Boor.
\newblock \emph{A Practical Guide to Splines}.
\newblock Springer-Verlag, New York, 1978.

\bibitem[Comanducci et~al.(2016)Comanducci, Magalh{\~{a}}es, Ubertini, and
  Cunha]{Comanducci.etal_2016}
G.~Comanducci, F.~Magalh{\~{a}}es, F.~Ubertini, and {\'{A}}.~Cunha.
\newblock On vibration-based damage detection by multivariate statistical
  techniques: Application to a long-span arch bridge.
\newblock \emph{Struct Health Monit}, 15\penalty0 (5):\penalty0 505--524, 2016.
\newblock \doi{10.1177/1475921716650630}.

\bibitem[Deraemaeker et~al.(2008)Deraemaeker, Reynders, Roeck, and
  Kullaa]{Deraemaeker.etal_2008}
A.~Deraemaeker, E.~Reynders, G.~D. Roeck, and J.~Kullaa.
\newblock Vibration-based structural health monitoring using output-only
  measurements under changing environment.
\newblock \emph{Mech Syst Signal Process}, 22\penalty0 (1):\penalty0 34--56,
  2008.
\newblock \doi{10.1016/j.ymssp.2007.07.004}.

\bibitem[Dierckx(1993)]{Dierckx_1993}
P.~Dierckx.
\newblock \emph{Curve and Surface Fitting with Splines}.
\newblock Oxford University Press, New York, 1993.
\newblock \doi{10.1093/oso/9780198534419.001.0001}.

\bibitem[Gertheiss et~al.(2024)Gertheiss, Rügamer, Liew, and
  Greven]{Gertheiss.etal_2024}
J.~Gertheiss, D.~Rügamer, B.~X.~W. Liew, and S.~Greven.
\newblock {Functional Data Analysis: An Introduction and Recent Developments}.
\newblock \emph{Biom J}, 66\penalty0 (7):\penalty0 e202300363, 2024.
\newblock \doi{10.1002/bimj.202300363}.

\bibitem[Greven and Scheipl(2017)]{Greven.Scheipl_2017}
S.~Greven and F.~Scheipl.
\newblock A general framework for functional regression modelling.
\newblock \emph{Stat Modelling}, 17\penalty0 (1-2):\penalty0 1--35, 2017.
\newblock \doi{10.1177/1471082X16681317}.

\bibitem[Han et~al.(2021)Han, Ma, Xu, and Liu]{Han.etal_2021}
Q.~Han, Q.~Ma, J.~Xu, and M.~Liu.
\newblock Structural health monitoring research under varying temperature
  condition: a review.
\newblock \emph{J Civ Struct Health Monit}, 11:\penalty0 149--173, 2021.
\newblock \doi{10.1007/s13349-020-00444-x}.

\bibitem[Hu et~al.(2017)Hu, Cunha, Caetano, Rohrmann, Said, and Teng]{Hu2017}
W.-H. Hu, {\'A}.~Cunha, E.~Caetano, R.~G. Rohrmann, S.~Said, and J.~Teng.
\newblock Comparison of different statistical approaches for removing
  environmental/operational effects for massive data continuously collected
  from footbridges.
\newblock \emph{Struct Control Health Monit}, 24\penalty0 (8):\penalty0
  e1955, 2017.
\newblock \doi{10.1002/stc.1955}.

\bibitem[Knoth(2017)]{Knoth_2017}
S.~Knoth.
\newblock {ARL} numerics for {MEWMA} {C}harts.
\newblock \emph{J Qual Technol}, 49\penalty0 (1):\penalty0 78--89, 2017.
\newblock \doi{10.1080/00224065.2017.11918186}.

\bibitem[Knoth(2024)]{Knoth_2024}
S.~Knoth.
\newblock \emph{spc: Statistical Process Control -- Calculation of ARL and
  Other Control Chart Performance Measures}, 2024.
\newblock {R} package version 0.7.1.
\newblock \doi{10.32614/CRAN.package.spc}.

\bibitem[Lowry et~al.(1992)Lowry, Woodall, Champ, and Rigdon]{Lowry.etal_1992}
C.~A. Lowry, W.~H. Woodall, C.~W. Champ, and S.~E. Rigdon.
\newblock {A Multivariate Exponentially Weighted Moving Average Control Chart}.
\newblock \emph{Technometrics}, 34\penalty0 (1):\penalty0 46--53, 1992.
\newblock \doi{10.2307/1269551}.

\bibitem[Maes and Lombaert(2020)]{Maes.Lombaert_2020}
K.~Maes and G.~Lombaert.
\newblock {Monitoring data for railway bridge KW51 in Leuven, Belgium, before, during, and after retrofitting. Zenodo}, 2020.
\newblock \doi{10.5281/zenodo.3745914}.

\bibitem[Maes and Lombaert(2021)]{Maes.Lombaert_2021}
K.~Maes and G.~Lombaert.
\newblock {Monitoring data for railway bridge KW51 in Leuven, Belgium, before, during, and after retrofitting}.
\newblock \emph{J Bridge Eng}, 26\penalty0 (3):\penalty0 04721001, 2021.
\newblock \doi{10.1061/(ASCE)BE.1943-5592.0001668}.

\bibitem[Maes et~al.(2022)Maes, {Van Meerbeeck}, Reynders, and
  Lombaert]{Maes.etal_2022}
K.~Maes, L.~{Van Meerbeeck}, E.~Reynders, and G.~Lombaert.
\newblock {Validation of vibration-based structural health monitoring on
  retrofitted railway bridge KW51}.
\newblock \emph{Mech Syst Signal Process}, 165:\penalty0 108380, 2022.
\newblock \doi{10.1016/j.ymssp.2021.108380}.

\bibitem[Magalh{\~{a}}es et~al.(2012)Magalh{\~{a}}es, Cunha, and
  Caetano]{Magalhaes.etal_2012}
F.~Magalh{\~{a}}es, A.~Cunha, and E.~Caetano.
\newblock Vibration based structural health monitoring of an arch bridge: From
  automated {OMA} to damage detection.
\newblock \emph{Mech Syst Signal Process}, 28:\penalty0 212--228, 2012.
\newblock \doi{10.1016/j.ymssp.2011.06.011}.

\bibitem[Prabhu and Runger(1997)]{Prab:Rung:1997}
S.~S. Prabhu and G.~C. Runger.
\newblock Designing a multivariate {EWMA} control chart.
\newblock \emph{J Qual Technol}, 29\penalty0 (1):\penalty0 8--15, 1997.
\newblock \doi{10.1080/00224065.1997.11979720}.

\bibitem[Scheipl et~al.(2015)Scheipl, Staicu, and Greven]{Scheipl.etal_2015}
F.~Scheipl, A.-M. Staicu, and S.~Greven.
\newblock {Functional Additive Mixed Models}.
\newblock \emph{J Comput Graph Stat}, 24\penalty0 (2):\penalty0 477--501, 2015.
\newblock \doi{10.1080/10618600.2014.901914}.

\bibitem[Wittenberg et~al.(2024)Wittenberg, Knoth, and
  Gertheiss]{Wittenberg.etal_2024b}
P.~Wittenberg, S.~Knoth, and J.~Gertheiss.
\newblock {Structural Health Monitoring with Functional Data: Two Case
  Studies}.
\newblock \emph{Preprint, arXiv:2406.01262}, 2024.
\newblock \doi{10.48550/arXiv.2406.01262}.

\bibitem[Wittenberg et~al.(2025)Wittenberg, Neumann, Mendler, and
  Gertheiss]{Wittenberg.etal_2025}
P.~Wittenberg, L.~Neumann, A.~Mendler, and J.~Gertheiss.
\newblock {Covariate-adjusted functional data analysis for structural health
  monitoring}.
\newblock \emph{Data-Centric Eng}, 6:e27, 2025.
\newblock \doi{10.1017/dce.2025.18}.

\bibitem[Wood(2017)]{Wood_2017}
S.~N. Wood.
\newblock \emph{Generalized Additive Models: An Introduction with R}.
\newblock CRC Press, Boca Raton, 2nd. edition, 2017.
\newblock \doi{10.1201/9781315370279}.

\bibitem[Xia et~al.(2012)Xia, Chen, Weng, Ni, and Xu]{Xia.etal_2012}
Y.~Xia, B.~Chen, S.~Weng, Y.-Q. Ni, and Y.-L. Xu.
\newblock Temperature effect on vibration properties of civil structures: a
  literature review and case studies.
\newblock \emph{J Civil Struct Health Monit}, 2:\penalty0 29--46, 2012.
\newblock \doi{10.1007/s13349-011-0015-7}.

\bibitem[Yao et~al.(2005)Yao, Müller, and Wang]{Yao.etal_2005}
F.~Yao, H.-G. Müller, and J.-L. Wang.
\newblock {Functional Data Analysis for Sparse Longitudinal Data}.
\newblock \emph{J Am Stat Assoc}, 100\penalty0 (470):\penalty0 577--590, 2005.
\newblock \doi{10.1198/016214504000001745}.

\end{thebibliography}

\end{document}